# Um Precursor das Ciências da Complexidade no Século XIX

Osame Kinouchi*
Departamento de Física, FFCLRP, Universidade de São Paulo

**Abstract**

The sciences of complexity present some recurrent themes: the emergence of qualitatively new behaviors in dissipative systems out of equilibrium, the aparent tendency of complex system to lie at the border of phase transitions and bifurcation points, a historical dynamics which present punctuated equilibrium, a tentative of complementing Darwinian evolution with certain ideas of progress (understood as an increase of computational power) etc. Such themes, indeed, belong to a long scientific and philosophical tradition and, curiously, appear already in the work of Friedrich Engels at the 70s of the XIX century. So, the apparent novelty of the sciences of complexity seems to be not situated in its fundamental ideas, but in the use of mathematical and computational models for illustrate, test and develop such ideas. Since politicians as the presidential candidate Al Gore declared that the sciences of complexity have influenced strongly their worldview, perhaps it could be interesting to know better the ideas and the ideology related to the notion of complex adaptive systems.

As ciências da complexidade apresentam alguns temas recorrentes: a emergência de comportamentos qualitativamente novos em sistemas dissipativos fora do equilíbrio, a aparente tendência de sistemas complexos de situarem-se na borda de transições de fase e pontos de bifurcação, dinâmicas históricas que apresentam equilíbrio pontuado, uma tentativa de complementar ideias de evolução Darwiniana com certas ideias de progresso (aumento de capacidade computacional) etc. Tais temas, na verdade, pertencem a uma longa tradição científica e filosófica e, curiosamente, aparecem já na obra de Friedrich Engels na década de 70 do século XIX. Assim, a aparente novidade das ciências da complexidade parece se situar não tanto em suas ideias fundamentais, mas no uso de modelos matemáticos e computacionais para ilustrar, testar e desenvolver tais ideias. Uma vez que políticos tais como o candidato à presidência dos EUA Al Gore declarou que as idéias das ciências da complexidade influenciaram fortemente sua visão de mundo, talvez seja interessante conhecer melhor as ideias e a ideologia ligadas à noção de sistemas complexos adaptativos.

# 1. Introdução

*As expected, many of our current themes on complexity have not arisen de novo, but may have been around for a long time, often in unexpected places (Harold Morowitz, Complexity, 2, 7-8, 1997).*

Uma das ideias sugeridas pelas ciências da complexidade, em particular pelos estudos que exploram os paralelos entre evolução cultural e evolução biológica, é a de que conceitos ou ideias nunca são totalmente novos, sendo sempre fruto de um longo processo histórico [Haas, 1998]. Essa ideia, na verdade, também não é nova e, na medida em que for válida, não poderia sê-lo. Assim, apesar das assertivas dos artigos e livros de divulgação científica, a própria abordagem das ciências da complexidade não é nova mas possui uma longa história que, às vezes, seus próprios entusiastas ignoram.

Vou ater-me aqui à versão das ciências da complexidade mais próxima da perspectiva do *Santa Fe Institute* (SFI) e que pode ser conhecida, por exemplo, através da revista *Complexity*. O SFI é uma entidade acadêmica privada cujo objetivo é catalisar colaborações transdisciplinares no estudo dos assim chamados sistemas complexos adaptativos (SCA). Dificilmente poder-se-ia considerar o SFI como ideologicamente próximo, por exemplo, da esquerda acadêmica americana. Por uma grande ironia histórica, porém, o SFI poderia ser considerado hoje como um dos mais ativos centros de divulgação de uma visão de mundo bastante próxima do materialismo histórico-dialético desenvolvido por Marx e Engels. Particularmente, poderíamos reconhecer em Friedrich Engels um precursor, já no século XIX, da abordagem de ciências da complexidade à la *Santa Fe Institute*, abordagem que pretende ser uma das grandes novidades do século XXI.

Friedrich Engels (1820-1895) é personagem histórica interessante. Filho de um comerciante alemão bem sucedido, recebeu forte educação religiosa ligada ao pietismo, fundamentalismo cristão florescente em Barmen, sua cidade natal. Ao ser enviado para Bremen para continuar sua formação de homem de negócios, é influenciado pelo ambiente cosmopolita desta cidade. Toma contato com idéias de teólogos radicais tais como Strauss e Schleimacher. Com cerca de 21 anos, liga-se ao movimento dos Jovens Hegelianos, intelectuais desejosos de utilizar as idéias de Hegel na crítica à religião e política contemporâneas. Em 1844, encontra Karl Marx em Paris, dando início a uma amizade e colaboração duradouras, cujo marco é o Manifesto Comunista (1848). Foi ele, na verdade, quem chamou a atenção de Marx para a importância da Economia no entendimento dos processos históricos e sociais. Sem o apoio e ajuda financeira de Engels, muito provavelmente a obra magna de Karl Marx, *O Capital*, nunca teria visto a luz do dia.

De 1850 a 1870, Engels reside em Manchester, levando uma vida dúplice como empresário e ativista comunista. Participa regularmente da caçada à raposa de Cheshire; é membro proeminente de dois clubes prósperos, o Albert Club e o Schiller Institute, do qual chegou a ser diretor [McLellan, 1979]. No seu círculo de amigos contam-se o jurista Samuel Moore e Karl Schorlemmer, titular da primeira cátedra inglesa de Química Orgânica. Ao mesmo tempo, vive um relacionamento particularmente feliz com uma jovem irlandesa de origem operária, Mary Burns, e prossegue suas intensas atividades de jornalista revolucionário e conselheiro de partidos socialistas. Na década de 70, já em Londres, gasta a maior parte de seu tempo estudando ciências naturais e Matemática, interesse que transparece no livro *Anti-Dühring* (1878), polêmica contra certo intelectual socialista, e em sua obra inacabada *Dialética da Natureza*, publicada postumamente apenas em 1927. Além dos trabalhos em colaboração com Marx, escreveu ainda *A Situação da Classe Operária na Inglaterra* (1844), *A Origem da Família, da Propriedade Privada e do Estado* (1884), *Ludwig Feuerbach e o Fim da Filosofia Clássica Alemã* (1888), *Do Socialismo Utópico ao Socialismo Científico* (1890).

Hoje, um século após sua morte (ocorrida em 1895), Engels é um autor marginalizado e esquecido. No decorrer do século XX, suas idéias foram criticadas e finalmente rejeitadas pelo liberalismo, pelo marxismo ocidental e pela esquerda romântica da década de 60. Basicamente, isso se deve ao fato de que o pensamento de Engels, simplificado e vulgarizado, foi adotado como doutrina oficial por Stalin. O fim da guerra fria e do socialismo burocrático, porém, talvez permita uma reavaliação mais serena do pensamento Engeliano. Parece agora ser possível reconhecer certas virtudes em suas ideias científicas e filosóficas sem que isto seja encarado como sintoma de um alinhamento político-ideológico estreito. Afinal, talvez seja mais fácil separar e recuperar Engels do Stalinismo do que fazer o mesmo com Nietzsche em relação ao Nazismo, tarefa na qual inúmeros filósofos contemporâneos têm-se dedicado.

Talvez a ideia mais original de Engels — e a que desperta maior rejeição — seja o que ele chamou de 'Dialética da Natureza'. Esta não é uma teoria ou concepção fechada, mas antes uma espécie de arcabouço conceitual que fornece certas ideias chaves e sugestões heurísticas a serem exploradas em pesquisas transdisciplinares. Mas não é apenas neste aspecto de paradigma científico alternativo, heurístico e um tanto frouxo, que as concepções de Engel se assemelham às das modernas ciências da complexidade. Os paralelos são bem mais fortes: por exemplo, Engels é fascinado pelas propriedades emergentes e transições de fase reveladas pela Termodinâmica e Física Estatística, e tenta aplicar estes conceitos (não apenas metaforicamente) à Biologia, Economia, Sociologia e História. Enfatiza que a evolução da matéria se dá historicamente por transições entre níveis de organização qualitativamente diferentes. Sua perspectiva dos processos naturais e sociais é dinâmica, sistêmica e ecológica: Engels define a Dialética como a ciência geral das mudanças e inter-conexões.

A ênfase em processos históricos é tão forte em Engels que ele chega a sugerir que as leis da Física não são fixas mas evoluem ao longo da história do Universo, uma conjectura que a atual Cosmologia tem explorado só recentemente. Sua concepção da dinâmica histórica, inspirada em Hegel, talvez constitua uma das primeiras formulações mais concretas do assim chamado equilíbrio pontuado. Para Engels, a estabilidade das configurações econômico-sociais dá lugar, de forma intermitente, a rápidas e dramáticas mudanças geradas a partir de uma dinâmica interna ao sistema. E essa dinâmica se deve não a um único fator causal[1], mas à co-evolução de diversos fatores. Por exemplo, Engels vê o processo de hominização como uma co-evolução, acelerada e retro-alimentada, entre características biológicas, sociais e culturais.

Metodologicamente, Engels é um teórico, crítico daquele empirismo estreito que não percebe que são as teorias que nos permitem definir e enxergar os 'fatos'. Engels simpatiza com idéias e modelos matemáticos simples (um de seus exemplos favoritos é a máquina de Carnot) que revelam o cerne de um processo por detrás da multidão de detalhes irrelevantes. Enfatiza a importância das formulações teóricas mais amplas e unificadoras, em contraste com o acumulo empírico de montanhas de dados colhidos sem critério teórico. Daí talvez seu entusiasmo pelas idéias de Darwin[2], paixão que lhe valeu severas críticas de outros marxistas por tentar aplicar à sociedade conceitos tomados à Biologia. É que Engels acreditava que as grandes mudanças paradigmáticas se dariam pela transferência de conceitos entre diferentes disciplinas e mesmo entre filosofia, ciências humanas e naturais. Finalmente, sua crença de que o avanço da Cosmologia produziria a superação das visões pessimistas derivadas da segunda lei da Termodinâmica (a chamada 'morte térmica do Universo'), o que curiosamente é verdade no contexto da Teoria da Inflação Cósmica e Multiverso, e sua insistência em postular uma tendência, inerente à matéria, à auto-organização e ao aumento de complexidade, pareceu a muitos um tipo de espiritualização encoberta.

Todas estas posições estão tão próximas daquelas das ciências da complexidade (ver, por exemplo, [Waldrop, 1992]) que certamente alguns leitores devem estar se perguntando se não estarei fazendo uma leitura demasiado forçada de Engels, projetando no pensador do século XIX conceitos e ideias próprias do final do século XX que lhe seriam estranhas. Assim, a seguinte estratégia será usada neste trabalho: colocarei lado a lado, com um mínimo de comentários, um grande número de citações de Engels e de autores que têm feito a divulgação das idéias de complexidade para o grande público. O uso de textos de divulgação científica em vez de artigos técnicos é proposital, pois nestes textos aparece mais claramente a discussão sobre a visão de mundo e as implicações filosóficas das ciências da complexidade, ou seja, seus aspectos ideológicos se apresentam de forma mais transparentes. Além disso, lembremos que Engels não é um filósofo ou cientista profissional, tendo atuado mais como jornalista e divulgador de ideias, de modo que a comparação entre os textos fica mais equilibrada. Em particular, darei atenção a autores ligados à perspectiva do *Santa Fe Institute* e da revista *Complexity*. Acredito que as citações sejam bastante extensas e numerosas para escapar da acusação de que seriam afirmações atípicas retiradas de seu contexto.

---

[1] A ênfase de Engels no caráter central dos fatores econômicos foi bastante matizada em seus escritos mais maduros.

[2] Após ler o Origem das Espécies, Marx escreveu à Engels que "este é o livro que provê a base de história natural para os nossos conceitos".

Se existem convergências, existem também divergências entre a visão de mundo Engeliana e as ênfases próprias das ciências da complexidade contemporâneas. No final do artigo, discuto brevemente as diferenças mais significativas, aquelas que porventura tiveram consequências perversas ao serem adotadas 'oficialmente' pelo socialismo tecno-burocrático. Finalmente, sugiro que a inesperada convergência entre a perspectiva de complexidade do SFI e a Dialética da Natureza Engeliana, desenvolvidas em contextos históricos e sociais tão diferentes, talvez seja um sinal de que tais ideias estejam ultrapassando o estágio de modismos transitórios, recorrentes, epidêmicos, para tornarem-se mais bem fundamentadas, enraizadas, culturalmente endêmicas. Afinal, o desenvolvimento de tais ideias ganha novo impulso e motivação em um mundo cada vez mais complexo, dinâmico, globalizado e inter-conectado. Que o mundo econômico-social constitui o meio ambiente seletivo na ecologia das ideias, influenciando fortemente sua disseminação, aceitação e estabelecimento, é também uma ideia Engeliana…

## 2 Convergências

Para Engels, a Dialética da Natureza consiste em uma visão de mundo ou perspectiva geral cuja utilidade terá que ser mostrada a posteriori, na medida em que frutificar em avanços científicos concretos. Essa visão de mundo, devida a Hegel (por sua vez influenciado por Aristoteles, por Heráclito e pela filosofia chinesa) é essencialmente dinâmica e sistêmica, enxergando o mundo mais como uma rede de processos do que como uma coleção de objetos, preferindo usar conceitos fluidos em vez de categorias fixas e bem delimitadas. Em particular, Engels sugere três temas heurísticos recorrentes ('leis da Dialética'):

1. Transformação da quantidade em qualidade: pequenas mudanças quantitativas podem induzir dramáticas mudanças qualitativas; a agregação de pequenas quantidades pode gerar propriedades qualitativamente novas.

2. Interpenetração dos opostos polares, identidade e luta dos contrários: é no conflito dinâmico entre opostos polares que se geram formas mais complexas.

3. Desenvolvimento através da contradição interna, ou negação da negação: a dinâmica interna dos sistemas complexos produz as próprias condições de sua superação, ou seja, a principal fonte de mudança é endógena. Para Engels, essa superação possui, no longo prazo e apesar de todos os retrocessos, uma tendência ascendente, progressiva, no sentido de emergência de novos níveis de organização da matéria.

Minha sugestão é que a abordagem de ciências da complexidade à lá SFI possui temas heurísticos similares, a saber:

1. Ênfase no estudo de propriedades coletivas emergentes e transições de fase em sistemas dinâmicos fora do equilíbrio.

2. Presumida tendência dos sistemas complexos adaptativos de se situarem na fronteira (pontos críticos, bifurcações etc.) entre dois comportamentos antagônicos. Exemplos de categorias bipolares: inatividade/atividade, ordem/desordem, periodicidade/Caos (borda do Caos), memória/inovação (janela de mutação), competição/cooperação etc.

3. Complexificação, ou aparente tendência 'espontânea' ao aumento de organização nos sistemas complexos devido à um processo de difusão no espaço (abstrato) de possíveis estruturas auto-organizadas.

É claro que as ciências da complexidade são muito mais ricas, teórica e experimentalmente, do que este conjunto de ideias genéricas. Seria uma caricatura exagerada dizer, parafraseando Lenin, que as ciências da complexidade equivalem à Dialética da Natureza mais computadores[3]. Minha intenção, porém, não é ficar venerando a memória de Engels mas apenas apresentá-lo como um precursor, no século XIX, de uma certa visão de mundo ou matriz paradigmática também compartilhada, parcialmente, pela Biologia morfológica de Darcy Thompson, a Biologia-Física de

---

[3] O objetivo de toda caricatura é revelar, pelo exagero, algo que passaria despercebido em um retrato mais fidedigno.

Alfred Lotka, a psicologia da Gestalt, a Cibernética, a Teoria de Sistemas, a Teoria de Catástrofes de Thom, a Sinergética de Haken, os sistemas auto-organizados de Prigogine, o Conexionismo (Redes Neurais Artificiais), a Criticalidade Auto-Organizada de Bak e os estudos de Sistemas Complexos Adaptativos. Embora muitas vezes constituindo modas científicas que floresceram e entraram em refluxo, sugiro que o que existe de comum a tais movimentos é um certo núcleo temático que também pode ser encontrado, agora de maneira mais bem fundamentada e permanente, na moderna Física Estatística e Teoria de Sistemas Dinâmicos.

Nas próximas seções, coloco trechos típicos de autores simpáticos à abordagem do SFI, seguidos por citações de Engels[4]. A idéia comum entre os trechos selecionado é colocada no título de cada subseção, existindo porém uma grande superposição de temas entre as mesmas de modo que cada seção ilumina e esclarece as outras, formando uma rede de referências cruzadas. É importante notar que os textos das próximas seções não são meu[5], mas dos autores citados no final de cada parágrafo. Espero que a simples superposição desses textos, aliada ao esforço do leitor em enxergar as inter-relações entre os mesmos, produza uma compreensão na forma de padrão gestáltico emergente que talvez seja mais eficaz que a discussão direta, explícita, linear e ordenada de cada um deles.

## 2.1 Metodologia: dialética na natureza, na sociedade, no pensamento e a ênfase na transdisciplinariedade

[O princípio de Universalidade] diz que, nos estados críticos existe um tipo de organização universal na qual os detalhes dos sistemas particulares deixam de ser importantes. Moléculas, átomos, magnetos ou spins — simplesmente não importa o que esteja interagindo. (...) Universalidade nos dá um novo entendimento de como sistemas aparentemente muito diferentes podem se comportar do mesmo jeito. Se você quer modelar algo como um magneto ou um fluido perto do ponto crítico, você não precisa se preocupar em representar acuradamente como cada componente interage com seus vizinhos. Qualquer modelo, não importa quão abstrato ou ridículo, servirá, na medida em que ele pertencer à mesma classe de universalidade do sistema original [Buchanan, 1997].

O comportamento de uma economia, uma companhia ou um ecossistema surge a partir das interações entre os indivíduos que os compõe. Sistemas cooperativos estão em todo lugar, sejam bandos de pássaros ou colônias de bactérias. E, de acordo com o princípio de Universalidade[6], a natureza exata dos elementos que fazem esses sistemas e como eles interagem é frequentemente irrelevante. Universalidade nos dá confiança, diz [o físico H. Eugene] Stanley, de que realmente podemos modelar e entender sistemas complexos como estes [Buchanan, 1997].

---

[4] As siglas AD, DN, LF e SUSC referem-se, respectivamente, aos livros *Anti-Düring*, *Dialética da Natureza*, *Ludwig Füerbach e o Fim da Filosofia Clássica Alemã* e *Do Socialismo Utópico ao Socialismo Científico*.

[5] Quando necessário, observações pessoais serão colocadas em colchetes.

*No presente trabalho, a Dialética é concebida como a ciência das leis mais gerais de todo movimento. Isto implica que suas leis precisam ser válidas tanto para o movimento na natureza e na história humana como o movimento do pensamento. Uma lei desse tipo pode ser reconhecidas em duas dessas três esferas, na verdade mesmo em todas as três, sem que o metafísico filisteu esteja consciente que a lei que ele veio a conhecer é uma e a mesma.*

*A Dialética tem sido até agora investigada mais de perto apenas por dois pensadores, Aristoteles e Hegel. Porém é precisamente a Dialética que constitui a mais importante forma de pensamento para a ciência natural contemporânea, pois apenas ela oferece a analogia, e portanto o método de explanação, para os processos evolucionários que ocorrem na natureza, inter-conexões em geral, e transições de um campo de investigação para outro* (DN, Old preface to Anti- Du̇ring).

*Para a Dialética, que focaliza as coisas e as suas imagens conceituais substancialmente nas suas conexões, na sua concatenação, na sua dinâmica, no seu processo de nascimento e caducidade, fenômenos como os expostos não são mais que outras tantas confirmações de seu modo genuíno de proceder. A natureza é a pedra de toque da Dialética, e as modernas ciências naturais oferecem-nos para esta prova uma acervo de dados extraordinariamente copioso e enriquecido a cada dia que passa, demonstrando com isso que a natureza se move, em última instância, pelos caminhos dialéticos e não pelas veredas metafísicas, que não se move na eterna monotonia de um ciclo constantemente repetido, mas percorre uma verdadeira história. Aqui é necessário citar, em primeiro lugar, Darwin, que, com a sua prova de que toda a natureza orgânica existente, plantas e animais, e entre eles, como é lógico, o homem, é o produto de um processo de desenvolvimento de milhões de anos, assestou na concepção metafísica da natureza o mais rude golpe* (SUSC).

*A Dialética, a chamada Dialética objetiva, impera em toda a Natureza; e a Dialética chamada subjetiva (o pensamento dialético) é unicamente o reflexo do movimento através de contradições que aparece em todas as partes da Natureza e que (num contínuo conflito entre os opostos e sua fusão final, formas superiores) condiciona a vida da Natureza* (DN, Notas).

---

[6] Buchanan usa a ideia de Universalidade em um sentido mais frouxo e amplo do que o usado em Física Estatística, onde este princípio é melhor fundamentado.

## 2.2 Metodologia: reducionismo, holismo, propriedades sistêmicas e universalidade

Assim, dos terremotos à evolução, a noção de universalidade reside atrás dessas teorias que estão adicionando uma dimensão extra no nosso entendimento do mundo. Mas as consequências dessa ideia podem se revelar ainda mais profundas. Por centenas de anos a Ciência tem seguido a noção de que as coisas podem sempre ser entendidas — e podem ser apenas entendidas — quebrando-as em partes menores, e conhecendo essas peças completamente. Sistemas no estado crítico — e eles parecem ser muito comuns — furam este princípio. Aspectos importantes de seus comportamentos têm pouco a ver com as propriedades detalhadas de seus componentes. A organização em um magneto, uma companhia ou um ecossistema não é devida às partículas, pessoas ou espécies que compõe estes sistemas [Buchanan, 1997].

C. Langton — Um mecanicista rigoroso vê todas as setas indo para cima, mostrando que a interação local causa alguma propriedade global, como a vida ou um ecossistema estável. E um vitalista rigoroso vê as setas apontando para baixo, indicando algum tipo de propriedade global mística que determina o comportamento das entidades do sistema. O que a ciência da complexidade lhe dá é a compreensão de que ambos são importantes, ligados num laço apertado e interminável de retroalimentação. O sistema inteiro representa um padrão dinâmico, com energia sendo dissipada através dele. Os vitalistas vão ficar desapontados se olharem este tipo de padrão para apoiar sua posição, porque, tirando-se a energia, toda a coisa desaba. Não há nada externo impulsionando o sistema; a dinâmica vem de dentro dele mesmo [Lewin, 1994].

*Vemos que a causa e o efeito são representações que somente regem, como tais, na sua aplicação ao caso concreto, mas que, examinando o caso concreto na sua concatenação com a imagem total do universo, se juntam e se diluem na ideia de uma trama universal de ações e reações, em que as causas e os efeitos mudam frequentemente de lugar e em que o que agora ou aqui é efeito adquire em seguida, aqui ou ali, o caráter de causa, e vice-versa.*

*A análise da Natureza nas suas diversas partes, a classificação dos diversos processos e objetos naturais em determinadas categorias, a pesquisa interna dos corpos orgânicos segundo as diversa estruturas anatômicas, foram outras tantas condições fundamentais a que obedeceram os gigantescos progressos realizados, durante os últimos quatrocentos anos, no conhecimento científico da Natureza. Esses métodos de investigação, porém, transmitiram-nos o hábito de focar as coisas e os*

*processos da natureza isoladamente, subtraídos à concatenação do grande todo; portanto, não na sua dinâmica, mas estaticamente; não como substancialmente variáveis, mas como consistências fixas; não na sua vida, mas na sua morte.*

*O método metafísico de pensar, por muito justificado e até necessário que seja em muitas zonas do pensamento, (...), tropeça sempre, cedo ou tarde, com uma barreira, ultrapassada a qual ele se converte em método unilateral, limitado, abstrato, e se perde em insolúveis contradições. Absorvido pelos objetos concretos, não consegue perceber a sua concatenação; preocupado com a sua existência, não atenta na sua origem nem na sua caducidade; obsedado pelas árvores, não consegue ver o bosque* (SUSC).

*Entre os homens de ciência, o movimento é sempre considerado (...) como movimento mecânico, como mudança de lugar. Isso é herança do século XVIII, pré-químico, e torna muito mais difícil a clara compreensão dos processos. O movimento, aplicado à matéria, é transformação em geral. Do mesmo equívoco provém também esta fúria de reduzir tudo a movimento mecânico, o que destrói o caráter específico das demais formas de movimento. É preciso não se interpretar, em face disso, que cada uma das formas superiores de movimento não esteja sempre, necessariamente, conectada a um movimento mecânico real (exterior ou molecular); (...) mas a presença dessas formas subsidiárias não esgota, em cada caso, a essência da forma principal. Algum dia, reduziremos experimentalmente, com toda segurança, o pensamento a movimentos moleculares e químicos, no cérebro; mas, por acaso, isso esgotará a essência do pensamento?* (DN, Notas).

**2.3 Metodologia: empirismo versus teorias gerais, e o papel dos modelos simples.**

Chris Langton —Estamos procurando as regras básicas que fundamentam todos estes sistemas [macro-evolução, morfogênese, ecossistemas, organização social, cognição], não apenas os detalhes de um deles (...). Chris, e outros como ele no Instituto [SFI] estão procurando princípios universais, regras fundamentais que dêem forma a todos os sistemas complexos adaptativos.

Perguntei a Stuart [Kauffman] se ele realmente está procurando verdades universais: — O que estou procurando é uma teoria profunda da ordem na Biologia. Se você considerar o mundo como John [Maynard Smith] o faz, então, nossa única opção como biólogos é a análise sistemática das máquinas basicamente acidentais e suas histórias evolutivas basicamente acidentais. Sei que não é só isso. Há algo mais. (...) Existem coisas que Darwin não tinha como saber. Uma delas era a auto-organização nos sistemas dinâmicos complexos. Se a nova ciência da complexidade tiver sucesso, ela intermediará o casamento entre a auto-organização e a seleção [natural]. Os biólogos acharão bastante difícil assimilar a idéia de auto-organização em sua atual visão de mundo[Lewin, 1994].

Para físicos e matemáticos, a teoria é o que conta. Experimentos meramente provêm um quadro [de testes] aproximado para a teoria. Em Biologia, esta tradição é revertida. Produção de dados é a prioridade, e qualquer teorização é fortemente adiada até que evidência experimental esteja disponível. O novo problema para ambos, contudo, é deixar estas tradições convergirem [Steimetz, 1997].

Um tipo importante de simulação nas Ciências Sociais é a modelagem baseada em agentes. Este tipo de simulação é caracterizado pela existência de muitos agentes que interagem entre si com pouca ou nenhuma direção central. As propriedades emergentes de um modelo baseado em agentes são portanto o resultado de processos *bottom-up* em vez de uma direção *top-down*. (...) O objetivo da modelagem baseada em agentes é enriquecer nosso entendimento dos processos fundamentais que podem aparecer em uma variedade de aplicações. É importante manter o modelo tão simples quanto possível (...). A complexidade dos modelos baseados em agentes deve residir nos resultados da simulação, e não nas assunções dos modelos [Axelrod, 1997].

*Um exemplo notável do que há de injustificado na pretensão segundo a qual a indução é a forma única ou ainda predominante da investigação científica, pode ser encontrada no terreno da Termodinâmica: a máquina a vapor constituía a demonstração mais assombrosa de que, do calor, é possível extrair-se movimento mecânico. Mas a verdade é que 100.000 máquinas a vapor não o demonstram melhor do que uma; criam apenas, para os físicos, a necessidade cada vez maior de explicar o fenômeno. Sadi Carnot foi o primeiro que se propôs a fazê-lo com seriedade. Mas não por meio da indução. Estudou a máquina a vapor, analisou-a, e verificou que o processo de seu funcionamento, aquilo que nela interessava, não se encontrava sobre uma forma simples mas encoberto por uma série de processos secundários; pôs de lado todas as circunstâncias estranhas ao processo essencial e construiu uma máquina a vapor ideal (ou máquina a gás), de construção por certo tão difícil como, por exemplo, uma linha ou superfície geométrica, mas que, de certa maneira, presta o mesmo serviço que essa abstrações matemáticas: apresentava o processo sob uma forma simples, independente, não adulterada. E topou, de repente, com o equivalente mecânico do calor... (DN, Notas).*

*No estudo da eletricidade [em meados do século XIX] impera uma confusa miscelânea de velhas experiências, ideias nem definitivamente confirmadas nem definidamente reprovadas, um inseguro tatear na obscuridade, um investigar e experimentar descoordenado, de muitos homens isolados, que atacam um território desconhecido, dispersos, como um bando de cavalos selvagens. (...) É principalmente*

*essa situação de abandono do estudo da eletricidade o que torna impossível, nesse período, a delineação de uma teoria geral; situação que dá origem, nesse terreno, ao domínio do empirismo unilateral, esse empirismo que, tanto quanto possível, proíbe-se a si mesmo de pensar e que, justamente por isso, não só pensa falsamente, como também não se coloca em condições de acompanhar fielmente os fatos ou de informar fielmente sobre os mesmos; e que, portanto, se converte no contrário do verdadeiro empirismo* (DN, Eletricidade (I)).

*Marx e eu fomos, sem dúvida alguma, os únicos que salvaram da filosofia idealista alemã a dialética consciente, incluindo-a na nossa concepção materialista da Natureza e da História. Mas uma concepção da História, a um tempo dialética e materialista, exige o conhecimento das matemáticas e das ciências naturais. Marx foi um consumado matemático (...). Ao fazer a recapitulação das matemáticas e ciências naturais, procurei convencer-me sobre uma série de pontos concretos — sobre o conjunto eu não tinha dúvidas — de que, na Natureza, se impõe, na confusão de das mutações sem número, as mesmas leis dialéticas do movimento que, também na história, presidem à trama aparentemente fortuita dos acontecimentos. (...) Leis essas primeiramente desenvolvidas por Hegel, mas sob uma forma que resultou mística, a qual o nosso esforço procurou tornar acessível ao espírito, em toda a sua simplicidade e valor universal* (AD, prefácio).

## 2.4 Transformação de quantidade em qualidade: propriedades emergentes e transições de fase.

Nessa altura, Normam Yoffee juntou-se ao grupo para uma breve visita. Antropólogo da Universidade do Arizona e especialista na dinâmica da formação do Estado, Norman descreveu a história das antigas civilizações da Mesopotâmia, o Iraque moderno. — A formação do Estado sempre acontece rapidamente. Os Estados s˜ao presumíveis e previsíveis. Chris Langton imediatamente reiterou o que tinha dito sobre as transições de fase em física e sua analogia com outros sistemas, inclusive as mudanças entre níveis diferentes de complexidade social. — Vejo tudo sob a ótica das transições de fase — admitiu. (...) Mas Chris tinha em mente algo mais do que simples analogia, algo mais que mera coincidência de padrão. – Talvez haja algo basicamente igual nos dois sistemas, de modo que os padrões sejam os mesmos, n˜ao importa quais os detalhes do sistema [Lewin, 1994].

Em níveis baixos de desenvolvimento tecnológico, podemos pensar que a economia esteja num estado estacionário correspondente ao estado estacionário de uma camada de fluido submetida a um aquecimento fraco. (...) Em níveis mais elevados de desenvolvimento tecnológico, ou de aquecimento, podemos contar com oscilações periódicas. De fato, ciclos econômicos aproximadamente periódicos foram observados. Em níveis ainda mais altos de desenvolvimento tecnológico, poderíamos ter uma superposição de duas ou três periodicidades diferentes, e os analistas econômicos viram tais coisas. Enfim, em níveis suficientemente altos de desenvolvimento, deveria haver uma economia turbulenta, com variações irregulares e uma dependência sensível das condições iniciais. Não deixa de ser razoável afirmar que atualmente vivemos numa tal economia. (...) Mas se tentarmos fazer uma análise mais quantitativa, topamos imediatamente com o fato de que os ciclos e outras flutuações da economia ocorrem sobre um fundo geral de crescimento. Há uma evolução histórica de sentido único que não podemos esquecer. De resto, os ciclos econômicos têm seu caráter histórico: cada um é diferente, não assistimos simplesmente a repetição monótona do mesmo fenômeno dinâmico. (...) Acho, porém, que nosso roteiro não é totalmente falso e que seu valor não é meramente metafórico. Por que? Porque não utilizamos certas propriedades subtilíssimas dos sistemas dinâmicos, mas, pelo contrário, robustos fatos de base. (...) Nosso roteiro, mesmo que tenha escasso valor quantitativo, pode, portanto, ser razoável qualitativamente [Ruelle, 1993].

*Na Natureza, de uma maneira fixada exatamente para cada caso individual, mudanças qualitativas podem somente ocorrer pela adição quantitativa ou quantitativa subtração de matéria ou movimento (a assim chamada energia). Se imaginarmos qualquer material não vivo sendo cortado em porções cada vez menores, de início nenhuma mudança qualitativa ocorre. Porém isto tem um limite: se conseguirmos, por exemplo por evaporação, obter as moléculas separadas em um estado livre, então é verdade que podemos usualmente dividi-las ainda mais, porém somente com uma completa mudança de qualidade. A molécula é decomposta em seus átomos individuais, os quais possuem propriedades muito diferentes daquelas da molécula. (...) os átomos livres de oxigênio são facilmente capazes de efetuar o que átomos de oxigênio atmosférico, ligados juntos na molécula, nunca podem alcançar (DN, Dialética).*

*Mas a molécula também é qualitativamente diferente da massa do corpo do qual ela pertence. Ela pode desenvolver movimentos independentes dessa massa enquanto esta última permanece aparentemente em repouso, ou seja, [apresentar] vibrações térmicas; e por meio de mudanças de posição e de conexão com moléculas vizinhas é possível mudar o corpo em um alótropo ou um estado diferente de agregação. Assim, vemos que a operação puramente quantitativa de divisão possui um limite no qual ela se torna em uma diferença qualitativa: o corpo consiste somente de moléculas, mas ele é algo essencialmente diferente da molécula, do mesmo modo que a última é diferente do átomo.*

*Uma intensidade de corrente mínima é requerida para fazer brilhar o fio de platina de uma lâmpada incandescente elétrica; e cada metal têm sua temperatura de incandescência e de fusão, cada líquido seus pontos definidos de congelamento e ebulição a uma dada pressão (...); finalmente, também cada gás possui seu ponto crítico no qual ele pode ser liquefeito por pressão e esfriamento. Em resumo, as assim chamadas constantes físicas são na sua maior parte nada mais que designações dos pontos nodais nos quais adição quantitativa ou subtração de movimento produz mudança qualitativa no estado do corpo considerado, nos quais, portanto, quantidade é transformada em qualidade.*

*Na Biologia, como na História da sociedade humana, a mesma lei vale a cada degrau, mas preferimos nos basear aqui em exemplos tirados das ciências exatas, uma vez que aqui as quantidades são acuradamente mensuráveis e podem ser seguidas. Provavelmente os mesmos cavalheiros que até agora descreveram a transformação de quantidade em qualidade como misticismo e transcendentalismo incompreensível agora irão declarar que ela é na verdade algo bastante auto-evidente, trivial e lugar-comum, a qual eles tem empregado há muito, de modo que não lhes foi colocado nada de novo. Mas ter formulado pela primeira vez na sua forma universalmente válida uma lei geral do desenvolvimento da natureza sociedade e pensamento, sempre permanecerá um ato de importância histórica* (DN, Dialética).

**2.5 Transições de fase e fenômenos sociais emergentes.**

Simulação é um modo de se fazer experiências de pensamento. Enquanto que as assunções podem ser simples, as consequências podem não ser óbvias. Os efeitos de larga escala de agentes interagindo localmente são chamados de 'propriedades emergentes' do sistema. Propriedades emergentes são frequentemente surpreendentes porque pode ser difícil antecipar todas as consequências mesmo de formas simples de interação [Axelrod, 1997].

O estudo de dilemas sociais provê insight numa questão central do comportamento: como cooperação global entre indivíduos confrontados com escolhas conflitantes pode ser assegurado. Estes avanços recentes mostram que comportamento cooperativo pode na verdade surgir espontaneamente em situações sociais, desde que os grupos sejam pequenos e diversos em composição e que seus constituintes possuam perspectivas [de interação] de longo prazo. Ainda mais importante, quando a cooperação aparece, isto acontece repentinamente e de forma imprevisível, após um longo período de estase [Glance & Huberman, 1994].

— Da interação dos componentes aqui em baixo surge uma espécie de propriedade global aqui em cima, algo que não poderia ter sido previsto a partir do que se sabe das partes componentes — continuou Chris Langton. — E a propriedade global, esse comportamento que surge, faz a retro-alimentação, influenciando o comportamento dos indivíduos aqui em baixo que o produziram. A ordem resultante de um sistema dinâmico complexo era como Chris a descrevia: propriedades globais brotando do comportamento gregário de indivíduos [Lewin, 1994].

*A História se faz ela mesma de tal maneira que o resultado final é sempre oriundo de conflitos entre muitas vontades individuais, cada uma das quais, por sua vez, é moldada por um conjunto de condições particulares de existência. Existem inumeráveis forças que se entrecruzam, uma série infinita de paralelogramos de forças que dão origem a uma resultante: o fato histórico. Este, por sua vez, pode ser considerado como o produto de uma força que, tomada em seu conjunto, trabalha inconscientemente e involuntariamente. Pois o desejo de cada indivíduo é frustrado pelo de outro, e o que resulta disso é algo que ninguém queria. Assim é que a História se realiza como se fosse um processo natural e está sujeita, também, essencialmente às mesmas leis de movimento.*

*Mas, do fato de que as diversas vontades individuais — cada uma das quais deseja aquilo a que a impelem a constituição física dos indivíduos e as circunstâncias externas (sejam pessoais ou da sociedade em geral que, em última instância, são econômicas) — não atinjam o que querem, mas se fundam numa média coletiva, numa resultante comum, não se deve concluir que o seu valor seja igual a zero. Pelo contrário, cada uma dessas vontades individuais contribui para a resultante e, nesta medida, está incluída nela. Eu pediria ao senhor que estudasse mais a fundo esta teoria nas suas fontes originais e não em fontes de segunda mão. Marx raramente escreveu alguma obra em que ela não tivesse seu papel, mas especialmente* O 18 Brumário de Louis Bonaparte *é um excelente exemplo de sua aplicação (carta de Engels a Konrad Schmidt, 5/8/1890).*

*Assim, por exemplo, em* O Capital de Marx, *toda a seção $4^a$, dedicada ao estudo da produção da mais-valia relativa ao âmbito da corporação, da divisão do trabalho, e da manufatura da maquinaria e da grande indústria, contém inúmeros casos de simples mudanças quantitativas que fazem com que se transforme a qualidade das coisas.(...) Temos, por exemplo, o fato de que a colaboração de muitas pessoas, a fusão de muitas forças numa só força total, cria, como diz Marx, uma nova potência de forças que se diferencia, de modo essencial, da soma das forças individuais associadas (AD, Cap. XII).*

*Somente depois de (...) fundamentar o fato de que não basta uma pequena soma qualquer de valor para que se possa converter em capital, mas que, para isso, um período todo de evolução e um ramo todo de produção deverão ultrapassar um*

*determinado limite mínimo, somente depois de tudo isso e em relação a estes fatos é que Marx adianta: 'Aqui, como nas ciências da natureza, se comprova a verdade da lei descoberta por Hegel em sua Lógica, segundo a qual, ao chegar a um determinado ponto, as mudanças meramente quantitativas se convertem em variações qualitativas'* (AD, Cap. XII).

## 2.6 A interpenetração dos opostos: Ordem versus desordem, estabilidade versus instabilidade e os estados críticos.

C. Langton —A velha visão do mundo da natureza era que ele pairava ao redor de equilíbrios simples. A ciência da complexidade diz que isso não é verdade. Os sistemas biológicos são dinâmicos, não facilmente previsíveis, e são criativos de muitas formas. (...) Na velha visão de equilíbrio do mundo, as ideias sobre mudança eram dominadas pela forma ação e reação. Era um mundo mecânico, aborrecidamente previsível ao máximo. Nessa espécie de mundo, você não podia ter avalanches de extinções e especiações de todas as magnitudes provocadas por uma mesma magnitude de mudança ambiental, por exemplo, como vemos nos modelos dinâmicos complexos [Lewin, 1994].

Você vê transições de fase a toda hora no mundo físico — disse Chris Langton. Você sabia que as membranas celulares estão apenas equilibradas entre o estado líquido e o sólido? Dê só um leve puxão, (...) deixe que uma única molécula de proteína se ligue a um receptor na membrana, e você poderá produzir grandes mudanças, mudanças biologicamente úteis. Perguntei se ele estava dizendo que as membranas biológicas estão no limite do caos, e não por acidente. — Estou. Estou dizendo que o limite do caos é onde a informação põe o pé na porta do mundo físico, onde ela exerce controle sobre a energia. Estar no ponto de transição entre a ordem e o caos não somente dá a você um controle apurado — pequeno estímulo/grande mudança – mas também permite que o processamento de informação se torne parte importante da dinâmica do sistema [Lewin, 1994].

*No organismo vivente, assistimos a um incessante movimento de todas as suas menores partículas, assim como de seus órgãos principais, donde resulta um continuado equilíbrio do organismo na sua totalidade, durante o período normal de vida e que, no entanto, sempre permanece em movimento, a vivente unidade de movimento e equilíbrio* (DN, Notas).

*O equilíbrio é inseparável do movimento (...). A possibilidade de um corpo ficar em equilíbrio relativo, a possibilidade de estados temporários de equilíbrio, é a condição essencial de diferenciação da matéria, e portanto, da vida. (...) Na superfície do Sol há um eterno movimento e inquietude, dissociação. Na Lua, parece prevalecer exclusivamente o equilíbrio, sem movimento relativo algum. Na Terra, o*

*movimento diferenciou-se, tendo-se estabelecido o intercâmbio entre movimento e equilíbrio: o movimento individual tende para o equilíbrio e o movimento, em seu conjunto, destrói mais uma vez o equilíbrio individual. (...) Todo equilíbrio é apenas temporário e relativo*. (DN, Notas).

**2.7 A interpenetração dos opostos: Memória versus mutação, tradição versus inovação e a evolução para a borda do caos.**

[No meu modelo computacional ecológico] se eu aumentar a taxa de mutação, o sistema deve se tornar caótico e extinguir-se. A uma taxa mais baixa, possivelmente não acontecerá nada de interessante. Entre essas duas velocidades, deveria produzir-se uma rica ecologia - disse-me Tom Ray a respeito de seu sistema Tierra [Lewin, 1994].

[No modelo de quasi-espécies de evolução molecular] o processo Darwiniano de organização fora do equilíbrio apresenta um paralelo claro com transições de fase ordem/desordem. No nosso caso, um valor de $q$ muito pequeno [$q \to 0$, alta fidelidade de cópia genética] leva a um único tipo de molécula (uma população viral uniforme), enquanto que altas taxas de erro ($q \to 1$) leva a um conjunto de moléculas totalmente aleatórias sem qualquer identidade biológica. (...) Observa-se uma transição abrupta para um certo valor de $q$, conhecida como catástrofe de erro. (...) Evidências experimentais mostram claramente que os retro-vírus estão tipicamente auto-organizados muito perto da catástrofe de erro. Neste sentido, o espectro largo de mutantes faz com que a otimização evolucionária se torne mais rápida [Solé et al., 1996].

*A teoria da evolução demonstra, tendo por base a simples célula, como cada progresso no sentido de uma planta mais complexa, por um lado, e no sentido do homem, por um outro, obedece à um contínuo conflito entre herança e adaptação. Em face disso, fica evidente como são pouco aplicáveis a tais formas de evolução categorias tais como positivo e negativo. Pode-se conceber a herança como algo positivo, conservador; e a adaptação como o lado negativo, que destrói continuamente as qualidades herdadas; mas igualmente se pode considerar a adaptação como sendo uma atividade criadora, positiva, e a herança como atividade resistente, passiva, negativa. (...) A teoria Darwiniana é a prova prática da íntima conexão entre acaso e necessidade conforme defendida por Hegel* (DN, Notas).

**2.8 A interpenetração dos opostos: Competição versus cooperação e a cooperação competitiva.**

A noção de se usar ecossistemas como uma metáfora para sistemas econômicos pode parecer bizarra. Afinal, a companhia ideal tem sido há muito pensada como uma máquina suavemente funcionando e sendo conduzida à objetivos específicos sob a direção de um onisciente, onipotente funcionário executivo chefe (CEO). A metáfora de companhias como espécies — alimentando-se do dinheiro dos consumidores e interagindo como em um ecossistema — traz algumas mudanças importantes. Primeiro, CEOs terão que se acostumar a pensar suas companhias não como máquinas mas mais como organismos vivendo em comunidades, o que muda a natureza de suas visões econômicas. Segundo, CEOs terão que perceber que têm muito menos controle sobre o destino de suas companhias do que gostariam de acreditar. Esta mudança no modo que líderes de negócios vêm seu mundo leva a um paralelo notável com mudanças recentes no pensamento dos ecologistas. Basicamente, é um afastamento da visão que encara o mundo como simples, previsível e rumando para o equilíbrio; é um reconhecimento que o mundo é complexo, imprevisível e está longe do equilíbrio. É também uma superação da visão de que a competição cabeça-a-cabeça é a força fundamental que dá forma às comunidades ecológicas e de negócios. A maioria dos negócios têm sucesso se outros também são bem sucedidos. Competição é parte do quadro, é claro, mas longe de ser a única parte. Cooperação e construção de redes mutuamente benéficas são importantes também. Bradenburger e Nalebuff descrevem esta estratégia conjunta com o termo co-opetição, o qual é também o título de seu livro [Lewin, 1997].

*Antes de Darwin, o que era enfatizado por seus seguidores atuais era precisamente o funcionamento cooperativo harmonioso da natureza orgânica, como o reino vegetal fornece aos animais alimento e oxigênio, e animais suprem plantas com adubo, amônia e ácido carbônico. Mas logo depois que as teorias de Darwin foram geralmente aceitas, essa mesma gente mudou de rumo e começou a ver em todo lugar nada mais que competição. Ambas as visões são justificadas dentro de certos limites, porém ambas são igualmente unilaterais e estreitas. A interação de corpos na natureza não-vivente inclui ambos harmonia e choques; em seres vivos, tanto cooperação consciente e inconsciente como consciente e inconsciente competição. Por conseguinte, no que respeita à Natureza, não é aceitável arvorar apenas a bandeira unilateral da luta. É também inteiramente pueril pretender resumir toda a múltipla riqueza da evolução histórica e complexidade na magra e unilateral frase 'luta pela existência'* (DN, Notas).

**2.9 A interpenetração dos opostos: Acaso versus necessidade e o caos determinista.**

À primeira vista, o determinismo Laplaciano não reserva nenhum lugar ao acaso: se lanço ao ar uma moeda, as leis da Mecânica Clássica determinam, em princípio, com certeza, se ela cairá cara ou coroa. Como o acaso e as probabilidades, na prática, desempenham um papel importante em nossa compreensão da Natureza, podemos ser tentados em rejeitar o determinismo. De fato, como veremos, o dilema acaso/ determinismo é amplamente um falso problema.

Em primeiro lugar, não há incompatibilidade lógica entre acaso e determinismo, já que o estado de um sistema no instante inicial, em vez de fixado de maneira precisa, pode ser disposto conforme certa lei de acaso. Se assim for, a qualquer outro instante, o sistema terá, também uma distribuição ao acaso, e essa distribuição poderá ser deduzida da distribuição do momento inicial, graças às leis da Mecânica. Na prática, o estado de um sistema no instante inicial nunca é conhecido com uma precisão perfeita, ou seja, sempre se admite um pouquinho de acaso no estado inicial do sistema. Veremos que esse pouquinho de acaso no instante inicial pode proporcionar muito acaso (ou muita indeterminação) num momento ulterior. Notamos assim que, na prática, o determinismo não exclui o acaso. No máximo pode-se dizer que — se se quiser — há como apresentar a Mecânica Clássica sem nunca falar de acaso. Veremos mais adiante que isso já não é verdade para a Mecânica Quântica. Assim, duas idealizações diferentes da realidade podem divergir muito do ponto de vista conceitual, mesmo se suas predições forem praticamente idênticas para uma ampla classe de fenômenos [Ruelle, 1993].

*Outra oposição que se acha enredada a metafísica é a de acaso e necessidade. (...) O senso comum e, com ele, a maioria dos homens de ciência, tratam a necessidade e o acaso como determinações que se excluem mutuamente e para sempre. Uma coisa, uma relação, um processo, ou é casual ou é necessário; mas não as duas coisas simultaneamente. Em vista disso, ambas existem lado a lado, na Natureza; esta contém toda classe de objetos e processos, entre os quais, uns são acidentais e outros necessários. O que interessa, portanto, é não confundir ambas as classes. Em posição contrária a essa opinião, está o determinismo, que se transferiu do materialismo francês para a ciência e que procura liquidar o acaso, desconhecendo-o. (...) O fato de que, esta noite, às quatro da madrugada, uma pulga tenha me mordido, e não às três ou às cinco, e justamente do lado direito do ombro e não na barriga da perna esquerda: todos esses fatos são produzidos por uma irrevogável concatenação de causa e efeito, por uma irremovível necessidade e, certamente, duma tal maneira, que a esfera gasosa da qual se originou o sistema solar estava já constituída de forma a que estes fatos teriam de se verificar assim e não de outro modo. Contrariando ambas as concepções, apareceu Hegel com as proposições, até então inauditas, segundo as quais (...) o acaso é necessário, que a necessidade se determina a si própria como acaso e que, por outro lado, o acaso é, talvez, uma necessidade absoluta. A ciência continuou ignorando, simplesmente, essas proposições (...) e teoricamente persistiu, por um lado, nas vacuidades mentais da metafísica de Wolff segundo a qual uma coisa ou é casual ou é necessária, mas não ambas ao mesmo tempo; ou então, nesse um pouco menos vazio determinismo*

*mecânico: o que nega o acaso, em geral por meio de palavras, para acabar reconhecendo-o na prática, em cada caso particular* (DN, Notas).

# 3  A negação da negação: dinâmica histórica, complexificação e as ideias de progresso

## 3.1 Aumento da complexidade na Biosfera

A ideia de emergência, tão antitética a grande parte da biologia moderna, é a principal mensagem da ciência da complexidade e seu papel no esclarecimento dos padrões da Natureza. A emergência da dinâmica auto-organizadora, a qual, se verdadeira, forçará a reformulação da teoria de Darwin. A emergência de uma criatividade na dinâmica dos sistemas complexos da natureza, a qual, se verdadeira, implica a existência de uma mão invisível que traz estabilidade do nível mais baixo até o mais alto na hierarquia ecológica, culminando na própria Gaia. E a emergência de um impulso inexorável para uma complexidade sempre maior, e processamento de informação na natureza, que, se verdadeiro, sugere a evolução de uma inteligência suficientemente poderosa para contemplar tudo que era inevitável. A vida, em todos os seus níveis, não é simplesmente uma coisa atrás da outra, mas o resultado de uma dinâmica interna fundamental comum [Lewin, 1994].

A visão de mundo Spenceriana é a de que maior complexidade é uma manifestação inevitável do sistema, e é impulsionada pela dinâmica interna dos sistemas complexos: heterogeneidade a partir de homogeneidade, ordem a partir do caos. A visão puramente Darwinista é que a complexidade é construída unicamente pela seleção natural, uma força cega, não direcional; e não há qualquer aumento inevitável na complexidade. A nova ciência da complexidade combina elementos de ambos: as forças internas e externas se aplicam, e uma maior complexidade pode ser esperada como uma propriedade fundamental dos sistemas dinâmicos complexos. Tais sistemas podem, através da seleção, conduzir-se à borda do caos, um processo constante de co-evolução, uma constante adaptação. Parte do fascínio da borda do caos é uma otimização da capacidade computacional, seja o sistema um autômato celular ou uma espécie biológica evoluindo com outras como parte de uma comunidade ecológica complexa. Na borda do caos pode-se construir cérebros maiores....[Lewin, 1994].

[As ideias de progresso e auto-organização histórica de forma alguma são aceitas atualmente pelos biólogos. Para comprovar isso, basta comparar declarações de biólogos que defendem posições ideológicas as mais divergentes, mas que concordam na ênfase do papel do acaso, e da falta de tendências gerais, na história.]

Sou hostil a todos os tipos de impulsos místicos em direção à maior complexidade — disse Richard Dawkins quando lhe perguntei se um aumento na complexidade computacional poderia ser considerado uma parte inevitável do processo evolutivo.

Michel Ruse — Você pode realmente dizer que um cérebro é melhor que uma concha? [Lewin, 1994].

Stephen Jay Gould — O progresso é uma idéia nociva, culturalmente contaminada, não testável e não operacional que deve ser substituída, se desejarmos compreender os padrões da história. (...) Com raízes que se estendem ao século XVII, o progresso como uma ética central alcançou o clímax no século XIX, com a revolução industrial e o expansionismo vitoriano. (…) Você não pode nos culpar por sermos fascinados pela consciência, é uma enorme interrupção na história da vida. Eu a vejo como um acidente peculiar, mas a maioria das pessoas não quer vê-la assim. Se você acreditar que há um aumento inexorável no tamanho do cérebro através da história evolutiva, então a consciência humana torna-se previsível, não um acidente excêntrico. A nossa visão da evolução é muito centrada no cérebro, um preconceito que distorce nossa percepção do verdadeiro padrão da história.

Edward Wilson [é uma excessão] — Centrados no cérebro — ele riu. — Esse não é o máximo, em termos do modo politicamente correto de raciocinar?... Preciso dizer mais? [Lewin, 1994].

R. Lewin — A maioria das espécies da Terra hoje são organismos unicelulares como no pré-Cambriano, e muito do resto são insetos. Isto não parece progresso inexorável em direção a uma maior complexidade, parece?

N. Packard — Estamos falando de sobrevivência. Sim, há um número incontável de nichos lá fora nos quais as espécies se dão muito bem com certos níveis de capacidade computacional. Mas onde a sobrevivência é contestada, na maioria das vezes, você verá um aumento. Pense nisto como uma constante exploração da utilidade de maior complexidade computacional na evolução. Às vezes ela traz uma vantagem, e isto é o que dá a você uma seta [no processo histórico].

Nessa altura, já tinha ficado claro que se Normam Packard estiver correto em sugerir que um aumento na capacidade computacional representa uma seta no processo evolucionário, muitos biólogos terão problemas em lidar com a mensagem que a nova ciência da complexidade pode estar lhes trazendo.

N. Packard — Não estou dizendo que cada organismo tem necessidade de se tornar mais complexo: o sistema como um todo se torna mais complexo. (...) As pessoas não gostam dela [da ideia de progresso] não por razões científicas, mas sociológicas.

Brian Goodwin — Suponha que você reedite o Big Bang. Quais são as probabilidades de conseguir a mesma tabela periódica de elementos naturais, as mesmas combinações de prótons, nêutrons e elétrons? Muito boas, ou assim sou levado a crer. Penso num retorno à explosão Cambriana do mesmo modo, não no mesmo grau, talvez, mas como uma imagem. Se houver atratores dinâmicos no espaço de possibilidades morfológicas, como acredito, então uma reedição da explosão Cambriana produziria um mundo muito mais parecido com o que conhecemos do que acredita Steven Jay Gould [que enfatiza o aspecto acidental da História]. Não seria idêntico ao que conhecemos, mas é possível que houvesse muitas semelhanças, fantasmas que reconheceríamos instantaneamente. Em outras palavras, a história evolutiva não seria uma coisa atrás da outra, mas, até certo ponto interessante, seria inevitável. Agora isto está se tornando uma espécie de refrão dos sistemas complexos adaptativos [Lewin, 1994].

*Que a matéria evolua a partir de si mesma o cérebro humano pensante é um puro acidente para uma visão mecanística, embora necessariamente determinado, passo a passo, onde ele acontece. Mas a verdade é que é da natureza da matéria avançar para a evolução de seres pensantes, de modo que isto necessariamente ocorre sempre que as condições para isto (não necessariamente idênticas em todos os lugares e tempos) estejam presentes* (DN, Ciência Natural e Filosofia).

*O movimento da matéria não é meramente o cru movimento mecânico, mera mudança de posição, ele é calor e luz, tensão elétrica e magnética, combinação e dissociação, química, vida e, finalmente, consciência* (DN, Notas).

*Não importa quão inumeráveis os seres orgânicos, também, que precisam surgir e desaparecer antes que animais com um cérebro capaz de pensamento desenvolvam-se no seu meio, e por um pequeno intervalo de tempo encontre condições adequadas para a vida, apenas para serem exterminados mais tarde sem misericórdia — nós temos a certeza de que a matéria permanece eternamente a mesma em todas as suas transformações, que nenhum de seus atributos pode jamais ser perdido, e portanto, também, que com a mesma necessidade férrea que ela irá exterminar na terra sua maior criação, a mente pensante, a matéria deverá em algum lugar e num outro tempo, produzi-la novamente* (DN, Prefácio).

## 3.2 Dinâmica histórica: transições entre atratores, revoluções e equilíbrio pontuado.

— Pessoal, já os vi antes — disse Chris [Langton]. — Vocês não eram arqueólogos. Eram biólogos. Eram linguistas. Economistas, físicos, todos os tipos de disciplinas. (...) Cada vez que um grupo de pessoas vem aqui para uma dessas conferências, há algum tipo de processo histórico em estudo. Sistemas evolutivos são assim. São processos singulares, de modo que não se podem compará-los diretamente a nada. Você gostaria de repetir o processo, ver o que acontece na segunda vez, na terceira, e assim por diante. Não pode, portanto é aí que nós entramos[com modelos evolutivos computacionais]. [Lewin, 1994].

Os sistemas mais complexos exibem o que os matemáticos chamam de atratores, estados nos quais o sistema, dependendo de suas propriedades, eventualmente se fixa. Imagine flutuar num mar revolto e perigoso, girando em redor de enseadas. Os redemoinhos se instalam, dependendo da topografia do fundo do mar e da corrente d'agua. Eventualmente, você será atraído para um destes vórtices. Você fica aí até que alguma perturbação maior ou mudança no fluxo d'agua o empurre para fora, e então será sugado por outro. Isso, cruamente, é como uma pessoa poderia considerar um sistema dinâmico com atratores múltiplos: tal como uma evolução cultural, com tribos, chiefdoms e Estados equivalentes a atratores. Esse mar mítico teria de ser arrumado de modo que a pobre pessoa que flutua fosse suscetível ao redemoinho um em primeiro lugar, ao qual se sucederia o redemoinho dois e assim por diante. Não haveria necessariamente uma progressão de um para dois, três e quatro. A história está cheia de exemplos de grupos sociais atingindo um nível mais alto de organização social, e então caindo [Lewin, 1994].

C. Langton — Aí estão todos esse bandos de caçadores lá fora, grupos de indivíduos, cada um capaz de fazer todas as tarefas do grupo. Cada um deles sabe ca¸car, reunir plantas comestíveis, fazer roupas, e assim por diante. Interagem entre si, especializam-se, e então... Bum!... transição de fase... tudo muda. Há um novo nível de organização social, um nível mais elevado de complexidade. Se você tem populações que interagem, e sua boa forma depende dessa interação, verá períodos de estase entremeados com períodos de mudança. Vemos isso em alguns de nosso modelos evolutivos, portanto, eu esperaria vê-lo aqui também.

Roger Lewin — Neste caso, a história não poderia ser descrita como simplesmente uma coisa depois da outra, poderia? [Lewin, 1994].

A pilha de areia vai de uma configuração a outra, não gradualmente, mas por meio de avalanches catastróficas. Devido à estatística de lei de potência, a maioria dos deslizamentos está associada com as grandes avalanches. As pequenas avalanches, embora sejam mais frequentes, não representam muita coisa. A evolução numa pilha de areia se dá em termos de revoluções, **como na visão da História de Karl Marx**. As coisas acontecem através de revoluções, n˜ao gradualmente, precisamente porque sistemas dinâmicos [complexos] estão afinados no estado crítico. Criticalidade auto-organizada é o modo da Natureza fazer transformações enormes em pequenas escalas de tempo [Bak, 1997].

*O materialismo moderno vê na História o processo de desenvolvimento da Humanidade, cujas leis dinâmicas é sua missão descobrir. (...) O materialismo moderno resume e compendia os novos progressos das ciências naturais, segundo os quais a Natureza tem também a sua história no tempo, e os mundos, assim como as espécies orgânicas que em condições propícias os habitam, nascem e morrem, e os ciclos, no grau em que são admissíveis, revestem-se de dimensões infinitamente mais grandiosas* (SUSC).

*A história do desenvolvimento da sociedade difere substancialmente, em um ponto, da história do desenvolvimento da natureza. Nesta — se prescindirmos da ação inversa exercida por sua vez pelos homens sobre a natureza —, os fatores que atuam uns sobre os outros e em cujo jogo se impõe a lei geral, são todos agentes inconscientes e cegos. (...) Em troca, na história da sociedade, os agentes são todos homens dotados de consciência, que atuam movidos pela reflexão ou pela paixão, perseguindo determinados fins. Porém esta distinção, por muito importante que seja para a investigação histórica, sobretudo de épocas e acontecimentos isolados, não altera em nada o fato de que o curso da história se rege por leis gerais de caráter interno.*

*Também aqui reina, na superfície e no conjunto, um aparente acaso; raras vezes acontece o que se deseja, e na maioria dos casos os muitos fins propostos se entrecruzam uns com outros e se contradizem, (...). As colisões entre as inumeráveis vontades e atos individuais criam no campo da história um estado de coisas muito análogo ao que impera na natureza inconsciente (...). Por isso, em conjunto, os acontecimentos históricos também parecem presididos pelo azar. Porém, ali onde na superfície das coisas parece reinar a casualidade, esta se encontra sempre governada por leis internas ocultas, e o que se trata é de descobrir estas leis.*

*Portanto, se se quer investigar as forças motrizes que (...) constituem os verdadeiros impulsos supremos da história, não havia que se fixar nos motivos de homens isolados, por muito relevantes que eles sejam, mas sim nos impulsos que movem a grandes massas, a povos em bloco, e, dentro de cada povo, a classes inteiras; e não momentaneamente, em explosões rápidas, como fogo de palha, senão em ações continuadas que se traduzem em grandes mudanças históricas* (LF).

## 3.2 Dinâmica histórica e cosmologia

O modo mais importante em que a cosmologia do século XX difere das cosmologias de Newton ou Aristoteles é que ela é baseada na compreensão de que o universo evoluiu dramaticamente ao longo do tempo. (...) O sucesso do modelo de Big Bang, junto com o fracasso da teoria do Estado Estacionário, nos deixa com um universo cujo estado presente precisa ser entendido como o resultado de processos físicos que ocorreram em tempos anteriores, quando ele era muito diferente. Assim, a cosmologia se tornou uma ciência histórica (...). A noção de evolução não desempenhou até agora um papel central similar na física de partículas elementares. Isto parece não ser muito natural, dado o relacionamento íntimo que está se desenvolvendo entre a física de partículas e a cosmologia. Certamente, precisaríamos nos perguntar o que a noção tradicional de que as leis da física representam verdades ahistóricas significa em um universo cuja origem nós podemos literalmente quase ver [Smolin, 1995].

*As leis eternas da Natureza se transformam, cada vez mais, em leis históricas. O fato de que a água se apresente no estado líquido entre $0^o$ e $100^o$ C é uma lei natural eterna, mas para que seja válida, é necessário haver: 1) água; 2) determinada temperatura; 3) pressão normal. Na Lua não há água, no Sol existem apenas seus elementos; para estes corpos celestes a lei, portanto, não existe. (...) No Sol, devido à sua elevada temperatura, as leis de combinação química dos elementos, não prevalecem ou só operam momentaneamente, nos limites da atmosfera solar, dissociando-se os compostos novamente, ao aproximarem-se do Sol. Nas nebulosas, talvez não existam sequer todos aqueles 65 elementos que conhecemos [no final do século XIX], os quais, por sua vez, poderão ser de natureza composta.*

*Por conseguinte, se quisermos falar de leis naturais gerais, uniformemente aplicáveis a todos os corpos — desde as nebulosas até o homem —, só nos restam a gravidade e talvez a forma mais geral da teoria referente à transformação da energia, isto é, a teoria mecânica do calor. Mesmo esta teoria, entretanto, se converte (com sua aplicação lógica geral a todos os fenômenos naturais) em uma representação histórica das sucessivas modificações que se verificam num sistema celeste, desde a sua origem até o seu desaparecimento; por conseguinte, numa história em que, a cada etapa, imperam leis diferentes, isto é, diferentes formas fenomênicas do mesmo movimento universal; e, sendo assim, não resta outra coisa, constante e universalmente válida, senão o movimento.* (DN, Notas).

## 4 Música, ordem e caos

Talvez seja uma observação trivial que o que chamamos de música não é nem uma seqüência totalmente aleatória de sons (ruído branco) nem uma seqüência excessivamente repetitiva e periódica. Se pensarmos no ruído branco como desordem e na seqüência periódica como ordem, a heurística de transição ordem-desordem e criticalidade sugere que a música interessante deverá estar relacionada com a borda dessa transição, ou seja, ruído pink (ou 1/f), que tem propriedades fractais (sistemas críticos também possuem propriedades fractais). Com efeito, diversos estudos tentam relacionar características estruturais musicais com ruído 1/f e fractalidade [Voss & Clarke, 1978; Gardner, 1978, Hsü & Hsü, 1991, Manaris et al., 2005]. Embora a questão seja bem mais ampla e mesmo polêmica, é interessante que tais ideias se coadunam com a proposta de que a complexidade surge na borda do caos, do SFI, e de que criatividade surge na interpenetração dos opostos polares (ordem/desordem), de Engels.

## 5 Divergências

Talvez a principal diferença entre a abordagem do SFI e o pensamento Engeliano é a de que os pesquisadores ligados a esse Instituto têm um interesse maior na auto-organização e processamento de informação em sistemas descentralizados: ecossistemas, mercados, sociedades de insetos, sistema imunológico, sistema nervoso, morfogênese etc. Ou seja, existe uma ênfase no paradigma de processamento de informação paralelo distribuído, sem controle central, na auto-organização de baixo para cima. Assim, a abordagem do *Santa Fe Institute,* ao enfatizar uma economia decentralizada e auto-organizada, é mais compatível com visões econômicas liberais[7] ou com ideias anarquistas de auto-gestão. Engels acreditava que os sistemas complexos anárquicos estavam sujeitos a cataclismos (*crashs* financeiros, ciclos econômicos depressivos e destrutivos etc.) e que o custo em vidas humanas desses processos era muito alto. A solução seria o controle científico do sistema econômico, o controle da complexidade. Um trecho típico de Engels sobre esta questão é seguinte, onde fica claro um pensamento não-linear:

*Em face da Natureza, como em face da Sociedade, o modo atual de produção só leva em conta o êxito inicial e mais palpável; e, no entanto, muita gente ainda se surpreende pelo fato de que as consequências remotas das atividades assim orientadas seja inteiramente diferentes e, quase sempre, contrárias ao objetivo visado; admiram-se de que a harmonia entre a oferta e a procura se transforme em seu oposto polar, como se verifica no transcurso de cada ciclo decenal da indústria e como também a Alemanha o experimentou, com um pequeno prelúdio, no krash; surpreendem-se de a propriedade privada, fundada no trabalho próprio, se desenvolver necessariamente no sentido da carência de propriedade ente os trabalhadores, enquanto que toda a propriedade se concentra, cada vez mais, nas mãos dos que não trabalham...* (DN, From ape to man).

---

[7] O pensamento liberal muitas vezes cai na concepção individualista da sociedade, ou seja, na de que a sociedade é simplesmente a superposição (linear) de comportamentos individuais. Certamente esta concepção não corresponde à perspectiva não-linear e emergentista compartilhada pelas ciências da complexidade e a dialética Engeliana.

De certa forma, a abordagem do SFI veio dar razão tanto à Adam Smith quanto à Engels. A mão invisível (auto-organização) na economia e na ecologia certamente existe e, no entanto, ela não é necessariamente benéfica para os seres humanos e para as espécies da Biosfera. A auto-organização rumo ao estado crítico, se otimiza a capacidade de adaptação e a criatividade do sistema, também o deixa suscetível a cataclismas econômicos, sociais e ecológicos (reações em cadeia destrutivas na Economia, *crashs* financeiros, extinções coletivas na Biosfera etc.). Assim, a ideia de uma "mão invisível" realmente presente no mercado, adaptativa, criativa, mas que não otimiza o bem estar coletivo e sendo na verdade perigosamente auto-destrutiva, poderia colocar em um novo patamar de discussão as perspectivas liberais e marxistas.

A abordagem de complexidade auto-organizada enfatizada pelo SFI, quando aplicada ao sistema macro-ecológico, também parece fundamentar de algum modo as especulações de Vernadsky [1926], Lovelock [1990] e Margulis [1997, 2000] a respeito da emergência de ciclos geofísico-químico-biológicos autocatalíticos e auto-regulados (Gaia) [Ghiralov, 1995]. Este tipo de visão sistêmica centrada na Biosfera, defendida pelos movimentos ambientalistas, contrasta com uma visão antropocêntrica em que a produção econômica é o valor primordial e preocupações ecológicas só teriam sentido na medida em que afetassem o bem-estar da Humanidade. A ênfase Engeliana na importância da maximização da produtividade econômica de alguma forma se refletiu nas políticas industriais do socialismo tecno-burocrático. No entanto, esta talvez seja uma leitura parcial e injusta de Engels. Os seguintes textos refletem tanto um certo antropocentrismo econômico quanto uma perspectiva ecológica mais ampla e cuidadosa:

*Os animais, como já indicamos, modificam, por meio de sua atividade, a natureza ambiente, da mesma forma (mas não no mesmo grau) que o homem; e essas transformações por eles produzidas em seu ambiente, atuam, por sua vez sobre os elementos causais, modificando-os. Isso porque, na Natureza, nada acontece isoladamente. Cada ser atua sobre o outro e vice-versa; é justamente porque esquecem esse movimento reflexo e essa influência recíproca, que os nossos naturalistas ficam impossibilitados de ver com clareza as coisas mais simples* (DN, From ape to man[8]).

*O animal apenas utiliza a Natureza, nela produzindo modificações somente por sua presença; o Homem a submete, pondo-a a serviço de seus fins determinados, imprimindo-lhe as modificações que julga necessárias, isto é, domina a Natureza. E esta é a diferença essencial e decisiva entre o Homem e os demais animais; e, por*

---

[8] O título completo do capítulo é *The part played by labour in the transition from ape to man*.

*outro lado, é o trabalho que determina essa diferença. Mas não nos regozijemos demasiadamente em face dessas vitórias humanas sobre a Natureza. A cada uma dessas vitórias, ela exerce a sua vingança. Cada uma delas, na verdade, produz, em primeiro lugar, certas consequências com que podemos contar; mas, em segundo e terceiro lugares, produz outras muito diferentes, não previstas, que quase sempre anulam essas primeiras consequências. Os homens que na Mesopotâmia, na Grécia, na Ásia Menor e noutras partes destruíram os bosques, para obter terra arável, não podiam imaginar que, dessa forma, estavam dando origem à atual desolação dessas terras ao despojá-las de seus bosques, isto é, dos centros de captação e acumulação de umidade. (...) Os propagadores da batata, na Europa, não sabiam que, por meio desse tubérculo, estavam difundindo a escrófula. E assim, somos a cada passo advertidos de que não podemos dominar a Natureza como um conquistador domina um povo estrangeiro, como alguém situado fora da Natureza; mas sim que lhe pertencemos, com a nossa carne, nosso sangue, nosso cérebro; que estamos no meio dela; e que todo o nosso domínio sobre ela consiste apenas na vantagem que levamos sobre os demais seres de poder chegar a conhecer suas leis e aplicá-las corretamente* (DN, From ape to man).

A coincidência com a perspectiva do SFI é literal:

Brian Arthur, economista ligado ao SFI — Uma dessas [visões de mundo] é o ponto de vista de equilíbrio que nós herdamos do Iluminismo — a ideia de que existe uma dualidade entre Humanidade e Natureza, e que existe um equilíbrio natural entre eles que é ótimo para o homem. E se você acredita nessa visão, então você pode falar sobre a otimização de políticas relativas a recursos naturais etc. [...] O outro ponto de vista é o da complexidade, no qual basicamente não existe dualidade entre Humanidade e Natureza. Nós somos parte da Natureza. Nós estamos no meio dela. Não existe divisão entre quem age e quem sofre as ações porque nós todos somos parte dessa rede interconectada. Se nós, como humanos, tentamos tomar a a¸ção em nosso favor sem conhecer como o sistema total irá se adaptar — por exemplo, derrubando a floresta tropical — nós colocamos em movimento uma sequência de eventos que irá voltar para nós e formar um padrão diferente para nós nos ajustarmos, tal como uma mudança climática global. (...) É uma visão de mundo que, década após década, está se tornando mais importante no Ocidente – tanto na Ciência como na cultura geral. Muito, muito devagar, têm havido uma mudança gradual de uma visão exploratória da Natureza — Humanidade versus Natureza — para uma abordagem que enfatiza a acomodação mútua entre Homem e Natureza. O que tem acontecido é que nós estamos começando a perder nossa inocência, nossa ingenuidade, a respeito de como o mundo funciona. Na medida em que começamos a entender os sistemas complexos, começamos a entender que fazemos parte de um mundo sempre mutável, interconectado, não linear e caleidoscópico. (...) Assim, qual é o papel do Santa Fe Institute nisso tudo? Certamente não o de se tornar outro *think tank* de políticas, embora sempre existam algumas pessoas que esperam que ele o seja. Não, a tarefa do Instituto é nos ajudar a olhar este rio sempre mutável e entender o que estamos vendo [Waldrop, 1992].

Entretanto, as modernas ciências da complexidade talvez coloquem um pouco mais de ênfase nos limites da predição e do controle dos sistemas complexos. Existe uma atitude de maior humildade frente à complexidade dos sistemas estudados. Atualmente, os pesquisadores se contentam com uma compreensão qualitativa da emergência de certas propriedades e, muitas vezes, chega-se a compreender porque a predição quantitativa não é possível mesmo em princípio (ver por exemplo, o impacto da ideia de caos determinista na Meteorologia, ou da ideia de criticalidade auto-organizada na previsão de terremotos, ambos justificando a imprevisibilidade dos grandes eventos). Nesse sentido, Engels era demasiadamente otimista, e esse otimismo humanista em relação ao poder da razão e consciência humanas acabou se cristalizando nas ideias de uma sociedade e economia planejadas. O comunismo Engeliano, com seu sonho de uma sociedade racionalmente planejada, seria o apogeu do Iluminismo.

Engels estava até certo ponto certo. Uma sociedade cientificamente planejada pode ser, em certos casos, mais eficiente economicamente. No entanto, no longo prazo, talvez a adaptabilidade seja um fator mais importante do que a simples eficiência. Lembremos que as grandes empresas, na sua tendência à burocratização e organização racional, parecem apresentar uma certa tendência à rigidez e ao envelhecimento: as empresas passam, o mercado fica. Engels queria uma economia gerida como uma grande empresa, lubrificada e racionalmente organizada. Lenin queria que o Partido fosse o CEO da sociedade. A atitude refratária de ambos frente a, por exemplo, ideias anarquistas, é de natureza tecnoburocrática: os trabalhadores não teriam capacidade técnico-administrativa para promover uma auto-gestão eficiente das empresas ou da sociedade.

A grande ironia histórica talvez seja a de que a própria Ciência sugere agora que a estratégia mais racional no longo prazo, tanto para sociedades como para empresas, é uma combinação dialética de organização e desorganização, controle e descentralização, planejamento e adaptação, racionalidade e aparente irracionalidade. O simples laissez-faire, a livre evolução das forças de mercado, levariam o sistema econômico global inevitavelmente para o estado crítico, com sua inevitável instabilidade (suscetibilidade a reações em cadeia e *crashs*) em forma de leis de potência. Para se evitar tal instabilidade seria necessário, no mínimo, a existência de mecanismos de controle de fluxos de capital em escala global. Tais controles talvez fossem capazes de produzir um sistema quasi-crítico: um sistema que possui a mesma flexibilidade de um sistema crítico sem apresentar necessariamente bolhas, *crashs* e reações em cadeia econômicas globais auto-destrutivas.

A questão da globalização de mercados envolve ainda outro problema pouco analisado: embora ecossistemas sejam vistos como paradigmas de sistemas distribuídos adaptativos e evolucionários, e sejam muitas vezes tomados como metáforas para se pensar os mercados, devemos lembrar que, em ecossistemas ricos e criativos (como as florestas tropicais e os recifes de coral), as espécies locais nunca estão em competição direta com espécies de ecossistemas similares: o puma americano e o tigre asiático não competem diretamente pelos mesmos recursos, e um deles inevitavelmente seria extinto se isso acontecesse. Ou seja, no jargão da Física Estatística, esses sistemas auto-regulados são "espacialmente extensos". Simplesmente não existe experiência prévia na Biosfera de sistemas fortemente interconectados, em que distâncias espaciais sejam abolidas. É bastante provável que sistemas como mercados globalizados impliquem em grande diminuição de diversidade ao mesmo tempo em que ocorre um aumento de velocidade de propagação de perturbações e consequente instabilidade.

Já a questão da distribuição desigual de renda (e, principalmente, de poder) na forma de leis de potência (Lei de Pareto) não é solúvel dentro do quadro capitalista, uma vez que os mecanismos concentradores detectados por Marx, Engels e Pareto (curiosamente chamado de Engels da burguesia), a saber, processos multiplicativos de acumulação de capital, são inerentes à esse tipo de sistema e só poderiam ser realmente superados em sistemas econômicos alternativos. Entretanto, é de se esperar que os sistemas mais adaptativos, quaisquer que sejam eles, continuem a ser sistemas descentralizados, conflitivos e dialéticos: sistemas pretensamente harmoniosos de convivência humana serão sempre autoritários, por privilegiarem o polo da ordem e reprimir o polo da desordem. Fica aqui a sugestão das ciências da complexidade e da Dialética: a sociedade humana mais robusta a longo prazo é aquela que se situa perto, mas não excessivamente perto, da borda do caos.

# Referências